\documentclass{article}
\usepackage{emulateapj,pstricks,apjfonts}

\def\1e{\mbox{1E0657--56}}

\def\chandra    {{\em Chandra}\/}
\def\xmm        {{\em XMM}\/}

\def\as         {$^{\prime\prime}$}

\def\lesssim{\mathrel{\hbox{\rlap{\hbox{\lower4pt\hbox{$\sim$}}}\hbox{$<$}}}}
\def\gtrsim{\mathrel{\hbox{\rlap{\hbox{\lower4pt\hbox{$\sim$}}}\hbox{$>$}}}}
\def\lax{\lesssim}

\makeatletter
\newenvironment{inlinefigure}{%
\def\@captype{figure}%
\noindent\begin{minipage}{0.999\linewidth}\begin{center}}
{\end{center}\end{minipage}\smallskip}
\makeatother

\begin{document}

\submitted{18 May 2002, astro-ph/0205333 (electronic preprint only)}

\lefthead{A1835 {\em CHANDRA}-{\em XMM} DISCREPANCY}
\righthead{MARKEVITCH}

\title{ON THE DISCREPANCY BETWEEN {\em CHANDRA}\/ AND {\em XMM}\/
TEMPERATURE PROFILES FOR A1835}

\author{Maxim Markevitch}

\affil{Harvard-Smithsonian Center for Astrophysics, 60 Garden St.,
Cambridge, MA 02138; maxim@head-cfa.harvard.edu}

\begin{abstract}

This short technical note addresses a large discrepancy between the
temperature profiles for the galaxy cluster A1835 derived by Schmidt et al.\
(2001) using \chandra\ and by Majerowicz et al.\ (2002) using \xmm. The
causes of this discrepancy may be instructive for the \chandra\ and \xmm\
cluster analyses in general. The observation used by Schmidt et
al.\ was affected by a mild background flare that could not be identified by
the usual technique. This flare biased upwards the measured temperatures at
large radii. The remaining discrepancy appears to be due to the \xmm\ PSF
scattering that was not taken into account in the published analyses. While
the \xmm\ PSF is narrow, the surface brightness of a typical cluster also
declines very steeply with radius. For the moderately distant, cooling flow
cluster A1835, about 1/3 of the observed \xmm\ brightness {\em at any
radius}\/ is due to the PSF scattering from the smaller radii. As a result,
the contamination from the bright cool cluster center biases low the
measured temperatures near the core, and in general, any temperature
gradients are underestimated.

\end{abstract}

\vspace{5mm}

\section{INTRODUCTION}

A1835 is a moderately distant ($z=0.25$), symmetric cluster with a strong
central X-ray brightness peak. Using \chandra\ data, Schmidt et al.\ (2001)
measured its temperature profile out to $r=4'$. It revealed a cool center
and a temperature increase to $\sim 13$ keV at $r=0.5-1\;h^{-1}_{50}$
Mpc. However, Majerowicz et al.\ (2002) derived a very different temperature
profile from the \xmm\ data. While they confirm the presence of the cool
central region, their outer temperature reaches only $\sim 7$ keV. The two
profiles in the common range of radii are shown in Fig.~\ref{fig:publ}. Such
a discrepancy at the qualitative level merits detailed investigation. I
first re-analyze the \chandra\ data.

\begin{inlinefigure}
\pspicture(0,14.4)(8.75,24.0)

\rput[tl]{0}(-0.3,23.9){\epsfxsize=9.2cm \epsfclipon
\epsffile{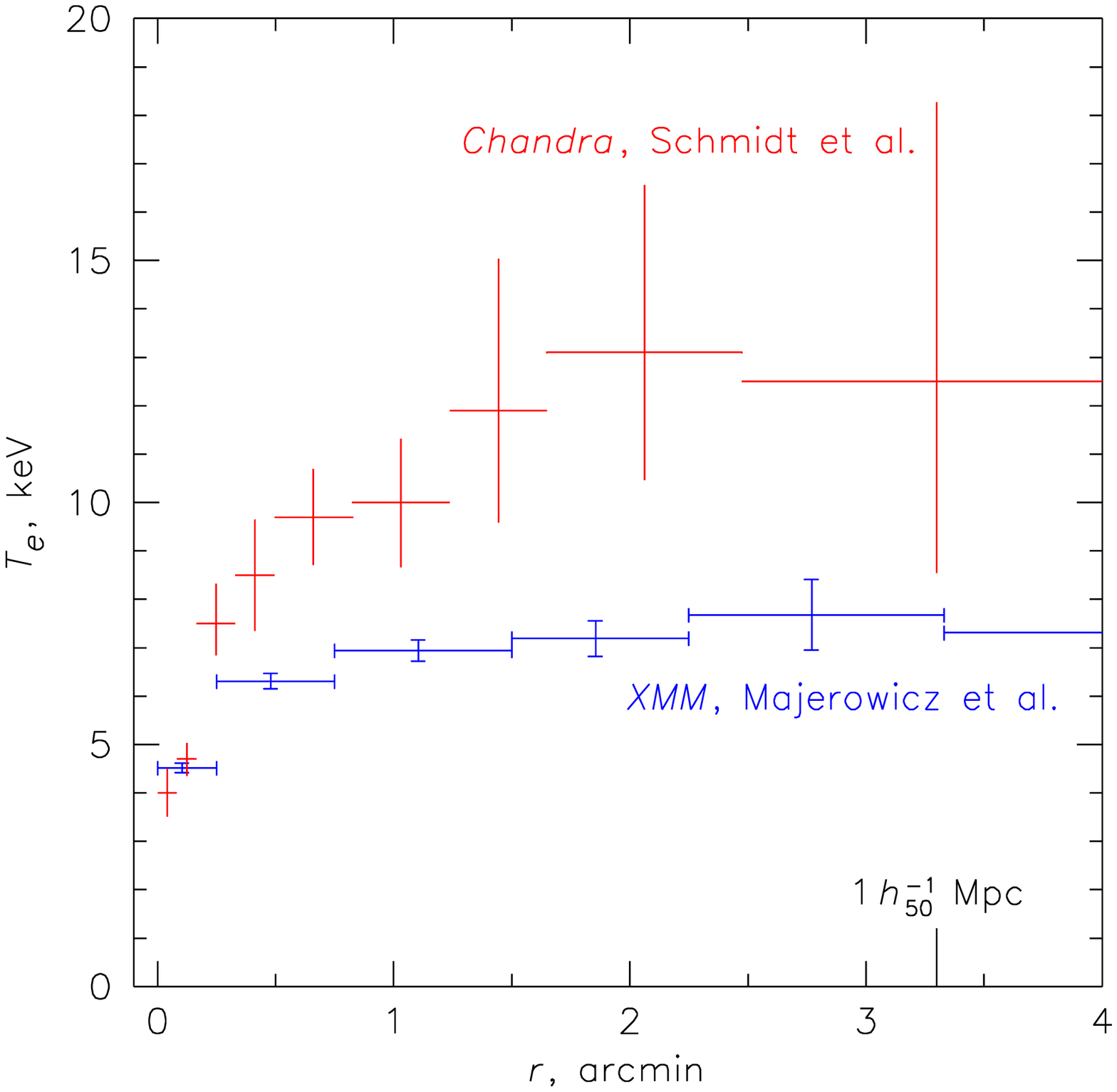}}

\rput[tl]{0}(0.0,15.2){
\begin{minipage}{8.75cm}
\small\parindent=3.5mm

\caption{Published \chandra\ (Schmidt et al.) and \xmm\ (Majerowicz et
  al.) temperature profiles for A1835. Errors are 90\%.}

\label{fig:publ}
\par
\end{minipage}
}

\endpspicture
\end{inlinefigure}

\section{{\it CHANDRA}\/ ANALYSIS} 

There were 2 \chandra\ observations of A1835, with the cluster in the ACIS
S3 chip in both. One (OBSID 495) with a 19 ks exposure was performed on
1999-12-11, and another (OBSID 496) with an 11 ks exposure was done on
2000-04-29. Schmidt et al.\ used only OBSID 495 for their spectral
analysis. They used 0.5--7 keV energy band for spectral fitting and the
public blank field dataset for background modeling. In OBSID 495, only one
backside-illuminated (BI) chip, S3, was used, and it was almost completely
covered by the cluster emission. This presents a difficulty for identifying
the background flares (see Markevitch 2001a) using the background light
curve. A light curve extracted from a region of the S3 chip far from the
cluster center does not vary beyond the normally acceptable $\pm20$\% range
from the mean (in agreement with the Schmidt et al.\ conclusions), although
due to the cluster contamination, one cannot compare that mean value to the
nominal background rate. In OBSID 496, one can also use S1 (another BI chip)
for flare detection; neither S1 nor S3 (its part far from the cluster
center) show any flares.

Since the quiescent background is known to vary on long and short timescales
at a few percent level, for each observation, I calculated the
normalizations of the corresponding blank-field datasets from the ratios of
count rates in the 10-12 keV band (which is free of the source emission and
of the most common variety of flares). As expected, such normalizations were
within a few \% of the respective exposure ratios.

ARFs and RMFs are calculated as described in, e.g., Markevitch \& Vikhlinin
(2001). The recently discovered low-energy ACIS quantum efficiency change
was corrected; it is not important for the present analysis. $N_H$ was fixed
at the Galactic value and abundances at 0.3 solar.

For OBSID 495, using the same background and energy band as Schmidt et al.,
I obtain a temperature profile very similar to theirs.

\begin{figure*}[t]
\pspicture(0,13.7)(18.5,23.4)

\rput[tl]{0}(-0.3,23.9){\epsfxsize=9.2cm \epsfclipon
\epsffile{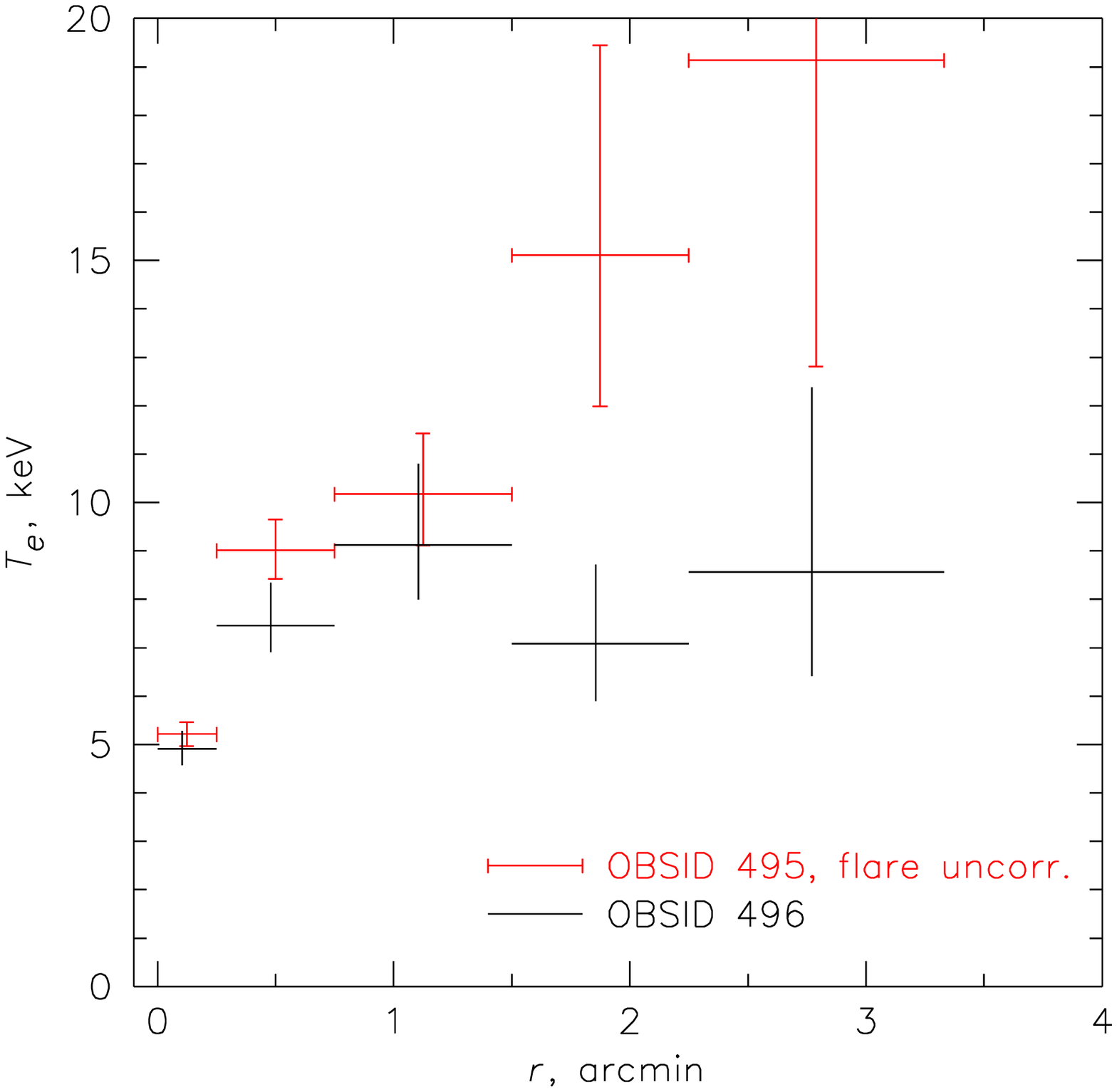}}

\rput[tl]{0}(-0.1,15.2){
\begin{minipage}{8.85cm}
\small\parindent=3.5mm

\caption{\chandra\ A1835 temperature profiles from observations OBSID 495
  (the one analyzed by Schmidt et al.) and OBSID 496, derived using the
  0.8--8 keV band. The soft Galactic background excess component was
  accounted for. Errors are 90\%.}

\label{fig:495uncorr}
\par
\end{minipage}
}

%

\rput[tl]{0}(9.35,23.9){\epsfxsize=9.2cm \epsfclipon
\epsffile{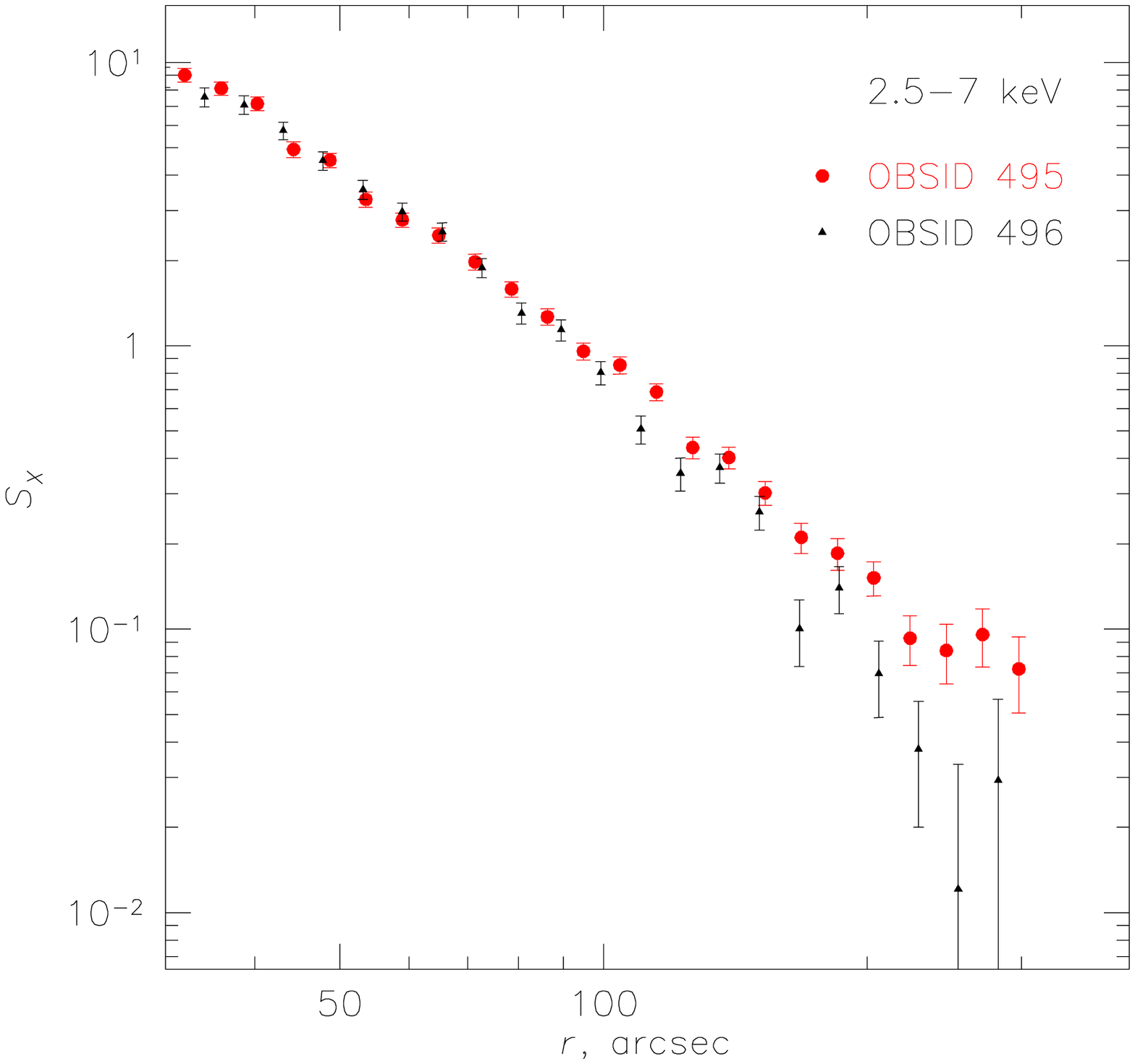}}

\rput[tl]{0}(9.65,15.2){
\begin{minipage}{8.85cm}
\small\parindent=3.5mm

\caption{\chandra\ radial brightness profiles from
  OBSIDs 495 and 496 in the 2.5--7 keV band (most sensitive to the background
  flares). The $S_X$ scale is arbitrary but the observations are normalized
  by their exposures. Errors are $1\sigma$. There is a clear background
  excess in OBSID 495.}

\label{fig:flareprof}
\par
\end{minipage}
}

\endpspicture
\end{figure*}

\subsection{Soft Galactic background excess}

However, as pointed out by Majerowicz et al., A1835 is projected onto the
Galactic North Polar Spur (Snowden et al.\ 1997) and thus has anomalously
high sky background at $E\lax 1$ keV. Since the public ACIS background
datasets represent the average high Galactic latitude background (see
Markevitch 2001b), this anomaly should be taken into account. To evaluate
this soft background excess, I extracted spectra from ACIS chips far from
the cluster, subtracted the nominal background, and fit the residuals by an
arbitrary model (I used low-temperature MEKAL with solar abundances). This
best-fit model (renormalized by the ratio of the solid angles) can then be
added to the fits in the interesting cluster regions. This properly takes
vignetting into account, assuming that the excess component is indeed
celestial in origin and does not vary on the ACIS FOV scale. This method was
used for \chandra\ analysis by Markevitch \& Vikhlinin (2001), and a similar
method (without the modeling step) was used for \xmm\ by Pratt et al.\
(2001) and Majerowicz et al.

For this procedure, a region of chips I2 and I3 at $r>10'$ (3 Mpc) was used
for OBSID 495, and the S1 chip was used for OBSID 496. I created subsets of
the blank-sky background datasets for the respective time periods (`B' for
OBSID 495 and `C' for OBSID 496) that included exactly the same blank fields
for S3 and those other chips (the next release of the public background
files will include such subsets). For the `C' period, such a subset
unfortunately is short and dominated by two pointings toward the Galactic
Spur. This resulted in {\em negative}\/ soft excess for OBSID 496, which was
fit by a {\em negative} model. In OBSID 495, there is a real excess above
the blank field background.

The best-fit soft excess (or deficit) was included in the cluster fits, and
a low energy cutoff of 0.8 keV was adopted to minimize the associated
uncertainty. The resulting radial profiles for the two observations are
shown in Fig.~\ref{fig:495uncorr}. The two are inconsistent, and OBSID 495
clearly has a problem.

\subsection{Flare background correction}

\begin{inlinefigure}
\pspicture(0,15.8)(18.5,24.1)

\rput[tl]{-90}(9.0,23.9){\epsfysize=9cm \epsfclipon
\epsffile[113 44 558 710]{flare_spec.ps}}

\rput[tl]{0}(-0.1,17.8){
\begin{minipage}{8.95cm}
\small\parindent=3.5mm

\caption{Spectrum from the region of the ACIS S3 chip outside $r=4.5'$
  ($1.4\,h^{-1}_{50}$ Mpc) from the cluster center, with the nominal
  background subtracted. There is a clear unphysical hard excess, which in
  the 2.5--7 keV band is consistent with the background flare model (shown
  by histogram; see text). The residuals at lower energies are due to cluster
  emission and the Galactic soft background excess.}

\label{fig:flarespec}
\par
\end{minipage}
}
\endpspicture
\end{inlinefigure}

The most obvious candidate for the problem in OBSID 495 is a mild background
flare, which are often observed in the BI chips. It may be too faint or too
constant to be identified in the light curve of the relatively short
observation (especially if one is using a region containing cluster emission
in addition to the background). Spectral shapes of the flare and quiescent
background components are such that the 2.5--7 keV band is most sensitive to
flares. Fig.~\ref{fig:flareprof} shows brightness profiles in this energy
band for the two observations, after subtracting the nominal background.
OBSID 495 shows a clear background excess.

\begin{figure*}[t]
\pspicture(0,14.3)(18.5,23.4)

\rput[tl]{0}(-0.3,23.9){\epsfxsize=9.2cm \epsfclipon
\epsffile{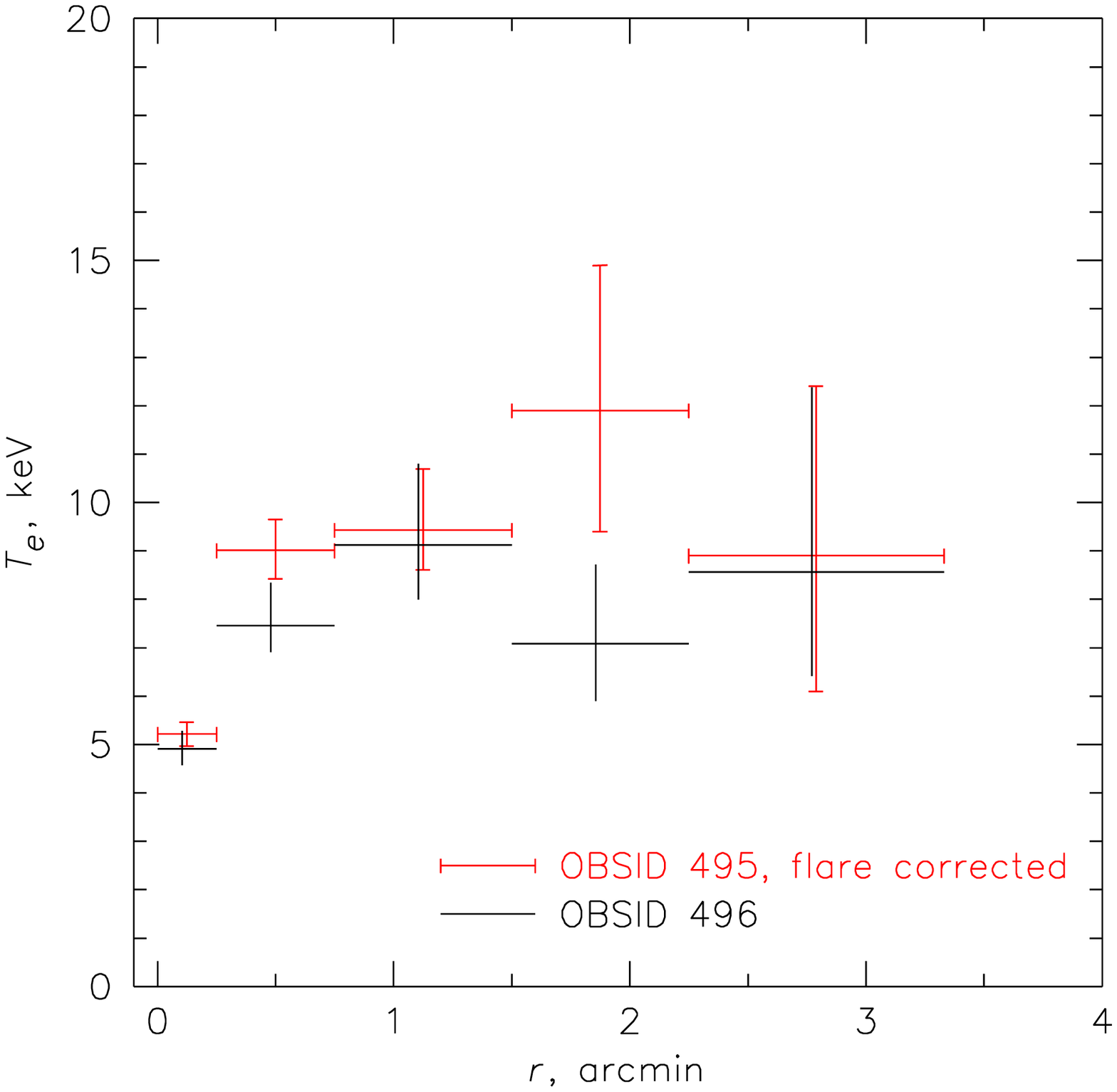}}

\rput[tl]{0}(-0.1,15.2){
\begin{minipage}{8.85cm}
\small\parindent=3.5mm

\caption{\chandra\ profiles from OBSIDs 495 and 496, now with the corrected 
  background flare component in OBSID 495. Errors are 90\%.}

\label{fig:495corr}
\par
\end{minipage}
}

%

\rput[tl]{0}(9.35,23.9){\epsfxsize=9.2cm \epsfclipon
\epsffile{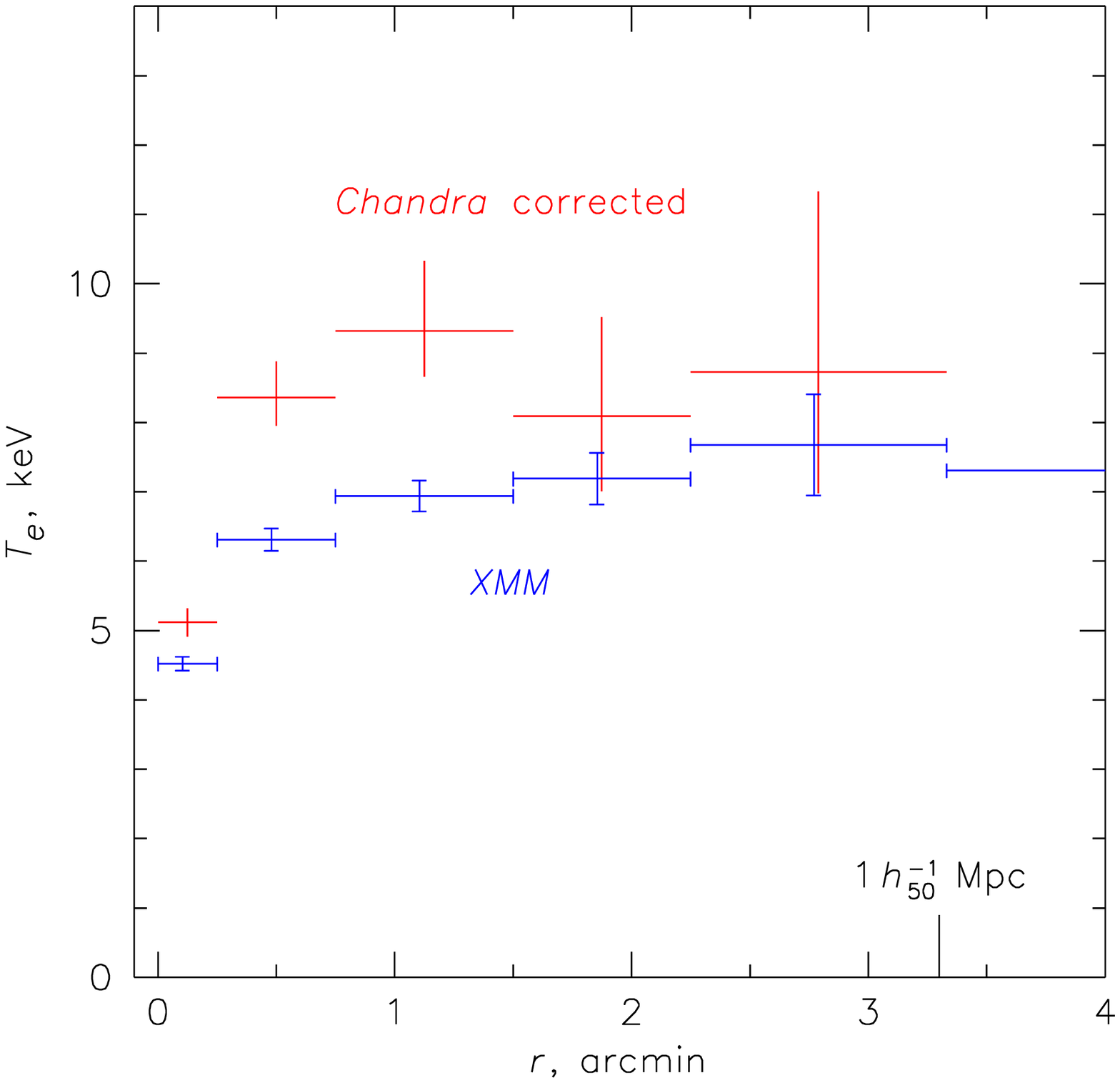}}

\rput[tl]{0}(9.65,15.2){
\begin{minipage}{8.85cm}
\small\parindent=3.5mm

\caption{The corrected \chandra\ temperature profile (average of 2
  observations) and the \xmm\ profile. Errors are 90\%.}

\label{fig:corrxmm}
\par
\end{minipage}
}

\endpspicture
\end{figure*}

Figure~\ref{fig:flarespec} shows a spectrum of this excess, extracted from the
region of the S3 chip outside $r=270''$ from the cluster center. If the
cluster brightness profile from OBSID 496 in Fig.~\ref{fig:flareprof} is
correct, the true cluster brightness in this band at these large radii is a
small fraction of the background excess.

There are at least two species of the background flares observed in the BI
chips. The most frequent one affects only the BI chips and is not seen in FI
chips. In OBSID 495, there are no flares in the FI chips (this can be
established confidently using the regions free of cluster emission covered
by those chips). Thus the putative flare in chip S3 should belong to this
species. Such flares appear to have the same spectral shape regardless of
their brightness, and are well-described by a power law with the photon
index $\gamma \approx -0.1$ and an exponential cutoff at $E\approx 5$ keV
(Markevitch et al.\ 2002a) without the application of the telescope or CCD
efficiency (`arf none' in {\sc xspec}). Figure~\ref{fig:flarespec} shows
this model with the normalization fitted to the spectrum in the 2.5--7 keV
band. The excess at $E>2.5$ keV is fully consistent with this model; the
residuals at lower energies should be due to the soft Galactic excess
(uncorrected for this exercise) and the cluster emission. At $E=3-5$ keV,
the flare brightness is 30--40\% of that of the nominal background.

Thus, the hard excess in OBSID 495 is consistent with a mild residual
background flare. We can assume (somewhat arbitrarily) that this component
is spatially uniform and subtract it from the cluster spectra, normalizing
it according to the solid angles. This method was used in Markevitch et al.\
(2002b). The resulting corrected temperature profile for OBSID 495 is shown
in Fig.~\ref{fig:495corr} along with the original profile for the unaffected
OBSID 496. A $\pm 40$\% uncertainty (90\% confidence) for the flare
component normalization is included in quadrature. The effect of the flare
correction is significant at $r>1'$. After the correction, the two \chandra\
observations become more or less consistent. (Note that the original Schmidt
et al.\ profile shown in Fig.~1 is consistent with the corrected profile;
the competing effects of the soft Galactic background excess and the hard
flare excess partially compensated for each other in that work.)

Figure~\ref{fig:corrxmm} overlays the corrected \chandra\ temperature
profile on the \xmm\ profile; there is still a significant discrepancy.

\section{{\it XMM}\/ PSF EFFECT}

\begin{inlinefigure}
\pspicture(0,14.3)(8.75,22.9)

\rput[tl]{0}(-0.3,23.9){\epsfxsize=9.2cm \epsfclipon
\epsffile{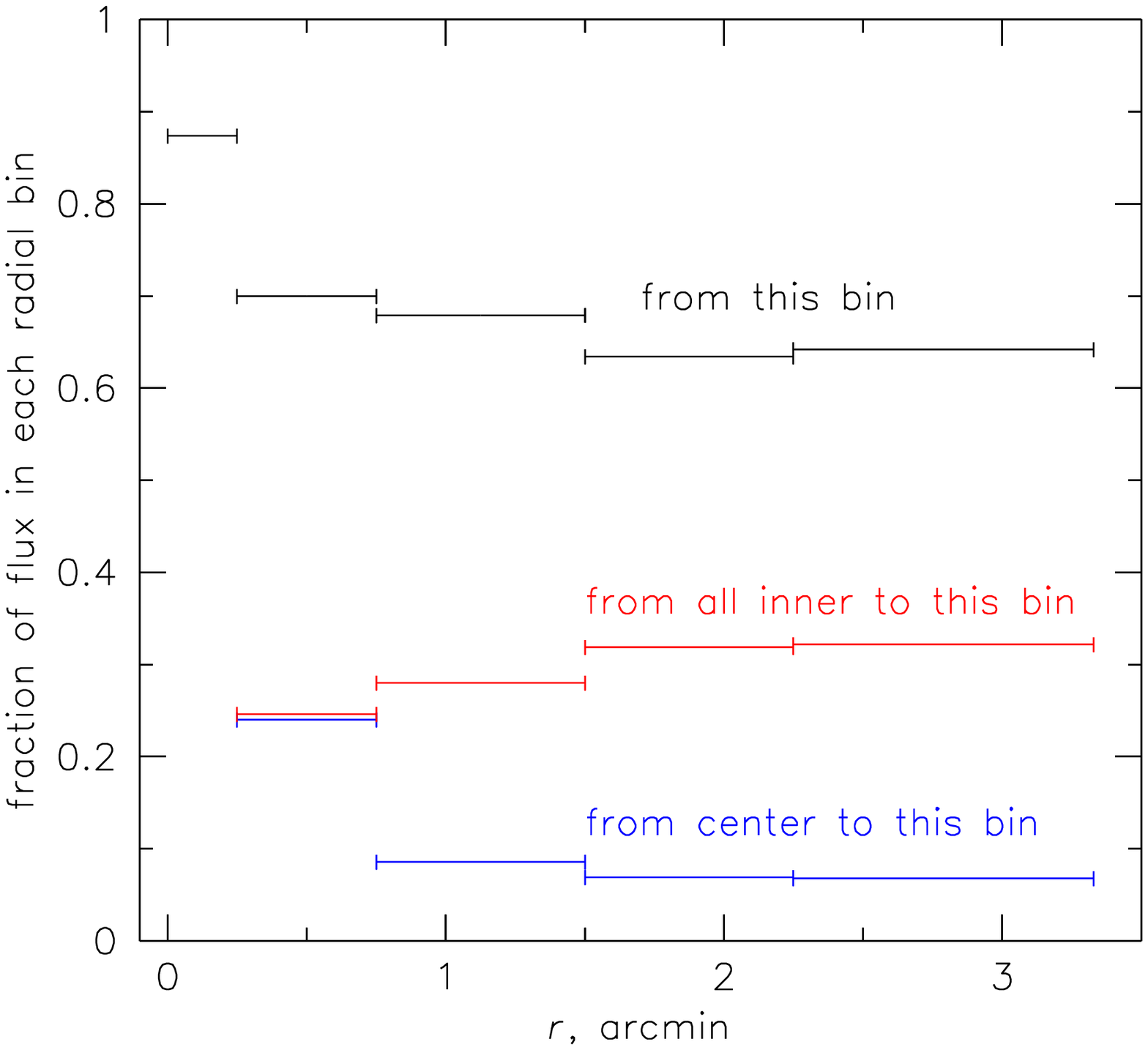}}

\rput[tl]{0}(0.0,15.2){
\begin{minipage}{8.75cm}
\small\parindent=3.5mm

\caption{Fraction of the observed flux in each  \xmm\ annulus
  originating from the same annulus in the sky, and fractions scattered by
  the telescope from the central circle and from all inner regions including
  the center.}

\label{fig:psf}
\par
\end{minipage}
}

\endpspicture
\end{inlinefigure}

The obvious candidate for the remaining discrepancy is the \xmm\ PSF, which
was not included in any currently published \xmm\ analyses. For a quick
estimate of the PSF effect, I use on-axis MOS1 PSF in the 0.75--2.25 keV
band presented in ``{\em
XMM-Newton}\/ Users' Handbook''.%
\footnote{http://heasarc.gsfc.nasa.gov/docs/xmm/uhb/xmm\_uhb.html, sections
``X-ray telescopes'', ``On-axis PSF''}
It can be described by the function $S_X \propto (1+r^2/r_c^2)^{-\alpha}$
with $r_c=3.5''$, $\alpha=1.36$. Ignoring its dependence on coordinate and
energy, one can estimate what fraction of the observed \xmm\ flux in each
region originates in the same region in the sky and how much is scattered
from other regions. I used the \chandra\ image of A1835 as a true brightness
model (the \chandra\ PSF can be ignored for the present purpose). The model
image was cut into annuli (same as used by Majerowicz et al.), each annulus
is convolved with the PSF, and contributions from each annulus in the sky
into each annulus in the image were calculated.

The result for the inner 5 annuli (those completely covered by \chandra) is
shown in Fig.~\ref{fig:psf}.  Even though the PSF 90\% encircled energy
radius is only 45--50\as, the cluster brightness profile declines so steeply
with radius that in each annulus, there is a $\sim 10$\% or higher
contribution from the central peak. Furthermore, every annulus has a $\sim
30$\% contamination from the inner regions, with the obvious effect of
smearing any temperature gradients in the cluster.

A simple {\sc xspec} simulation (using EPIC responses and the exposure time,
energy band and spectral binning given by Majerowicz et al.) shows that if
the temperatures from the \chandra\ profile are mixed in the proportion
shown in Fig.~\ref{fig:psf}, one can obtain single-temperature EPIC fits
very close to those reported. In particular, for the 2nd annulus, I obtain
6.4--6.8 keV compared to \chandra's 8 keV. For the 3rd annulus, I obtain
7.0--7.7 keV compared to \chandra's 9 keV.

\section{LESSONS}

1. When ACIS background is critical, one should ensure that the quiescent
background rate is consistent with the nominal rate (the one in the
background datasets), even if there are no obvious signs of background
flares in the background light curve. For ACIS BI chips, the 2.5--7 keV band
is best for this purpose because it is sensitive to the most common flare
species.  The next version of the public background datasets for the BI
chips will be filtered using this energy band (instead of the presently used
wide band).

2. The \xmm\ PSF has to be taken into account for the analysis of clusters
with peaked X-ray brightness profiles. (Monique Arnaud communicates that it
will be included in their forthcoming reanalysis of the A1835 data.)

Incorrect modeling of the \chandra\ background and disregard of the \xmm\
PSF appear to explain most if not all of the discrepancy between the
recently published A1835 temperature profiles.

\acknowledgments 

This analysis was supported by NASA contract NAS8-39073 and grant NAG5-9945.

\end{document}